\documentclass[aps,prd,preprint,tightenlines,nofootinbib]{revtex4}
\usepackage{epsfig}
\usepackage{citesort}
\usepackage{graphicx}
\usepackage{dcolumn}
\usepackage{bm}

\newcommand{\etal}{{\it et al.}}
\newcommand{\cn}{Collaboration}
\def\ibid{{\it ibid.}}
\def\plB#1,#2(#3){{\rm Phys.\ Lett.\ }{\bf B#1}, {\rm#2} {\rm(#3)}}
\def\prD#1,#2(#3){{\rm Phys.\ Rev.\ }{\bf D#1}, {\rm#2} {\rm(#3)}}
\def\prC#1,#2(#3){{\rm Phys.\ Rev.\ }{\bf C#1}, {\rm#2} {\rm(#3)}}
\def\mplA#1,#2(#3){{\rm Mod.\ Phys.\ Lett.\ }{\bf A#1}, {\rm#2} {\rm(#3)}}
\def\npB#1,#2(#3){{\rm Nucl.\ Phys.\ }{\bf B#1}, {\rm#2} {\rm(#3)}}
\def\epjC#1,#2(#3){{\rm Eur.\ Phys.\ J.\ }{\bf C#1}, {\rm#2} {\rm(#3)}}
\def\prl#1,#2(#3){{\rm Phys.\ Rev.\ Lett.\ }{\bf #1}, {\rm#2} {\rm(#3)}}
\def\jhep#1,#2(#3){{\rm JHEP\ }{\bf #1}, {\rm#2} {\rm(#3)}}
\begin{document}

\preprint{\tighten\vbox{\hbox{\hfil EFI 08-03}
                        \hbox{\hfil SUHEP 08-03}}}

\title
{\LARGE Decay Constants of Charged Pseudoscalar Mesons}

\author{Jonathan L. Rosner}
\affiliation{Enrico Fermi Institute, University of Chicago, Chicago, IL 60637}
\author{and Sheldon Stone}
\affiliation{Department of Physics, Syracuse University, Syracuse, NY 13244\\
\\}
\date{June 5, 2008}

\begin{abstract}
We review here the physics of purely leptonic decays of $\pi^\pm$, $K^\pm$, $D^{\pm}$, $D_s^\pm$, and $B^\pm$ pseudoscalar mesons.
The measured decay rates are related to the product of the relevant weak interaction based CKM matrix element of the constituent quarks and a strong interaction parameter related to the overlap of the quark and anti-quark wave-functions in the meson, called the decay constant $f_P$. The interplay between theory and experiment is different for each particle. Theoretical predictions that are necessary in the $B$ sector can be tested, for example, in the charm sector. One such measurement, that of $f_{D_{s}}$, differs from the most precise unquenched lattice calculation and may indicate the presence of new intermediate particles, or the theoretical prediction could be misleading. The lighter $\pi$ and $K$ mesons provide stringent comparisons due to the accuracy of both the measurements and the theoretical predictions. This review was prepared for the Particle Data Group's 2008 edition.
\end{abstract}
\maketitle

\begin{figure}[htb]
\centerline{\epsfxsize=3.0in \epsffile{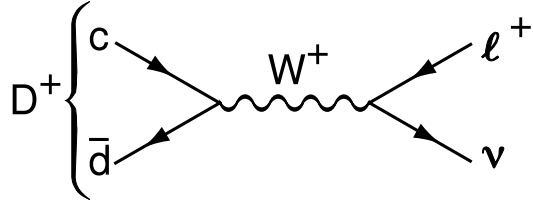}}
\caption{Annihilation process for pure $D^+$ leptonic decays in the
Standard Model.} \label{Ptoellnu}
\end{figure}

Charged mesons formed from a quark and anti-quark can decay to a
charged lepton pair when these objects annihilate via a virtual
$W^{\pm}$ boson. Fig.~\ref{Ptoellnu} illustrates this process for
the purely leptonic decay of a $D^+$ meson.

Similar quark-antiquark annihilations via a virtual $W^+$ ($W^-$) to
the $\ell^+ {\nu}$ ($\ell^- \bar{\nu}$) final states occur for the
$\pi^\pm$, $K^\pm$, $D_s^\pm$, and $B^\pm$ mesons. Let $P$ be any of
these pseudoscalar mesons.  To lowest order, the decay width is
\begin{equation}
\Gamma(P\to \ell\nu) = {{G_F^2}\over 8\pi}f_{P}^2\ m_{\ell}^2M_{P}
\left(1-{m_{\ell}^2\over M_{P}^2}\right)^2 \left|V_{q_1
q_2}\right|^2~. \label{equ_rate}
\end{equation}
Here $M_{P}$ is the $P$ mass, $m_{\ell}$ is the $\ell$
mass, $|V_{q_1 q_2}|$ is the Cabibbo-Kobayashi-Maskawa (CKM) matrix
element between the constituent quarks $q_1 \bar q_2$ in $P$, and
$G_F$ is the Fermi coupling constant. The parameter $f_P$ is the
decay constant, and is related to the wave-function overlap of the
quark and antiquark.

The decay $P^\pm$ starts with a spin-0 meson, and ends up with a
left-handed neutrino or right-handed antineutrino.  By angular
momentum conservation, the $\ell^\pm$ must then also be left-handed
or right-handed, respectively. In the $m_\ell = 0$ limit, the decay
is forbidden, and can only occur as a result of the finite $\ell$
mass.  This helicity suppression is the origin of the $m_\ell^2$
dependence of the decay width.

There is a complication in measuring purely leptonic decay rates.
The process $P\to \ell\nu\gamma$ is not simply a radiative
correction, although radiative corrections contribute. The $P$ can
make a transition to a virtual $P^*$, emitting a real photon, and
the $P^*$ decays into $\ell\nu$, avoiding helicity suppression. The
importance of this amplitude depends on the decaying particle and
the detection technique.  The $\ell\nu\gamma$ rate for a heavy
particle such as $B$ decaying into a light particle such as a muon
can be larger than the width without photon emission \cite{Bradcor}.
On the other hand, for decays into a $\tau^{\pm}$, the helicity
suppression is mostly broken and these effects appear to be small.

Measurements of purely leptonic decay branching fractions and
lifetimes allow an experimental determination of the product
$\left|V_{q_1 q_2}\right| f_{P}$. If the CKM element is well known
from other measurements, then $f_P$ can be well measured. If, on the
other hand, the CKM element is less well or poorly measured, having
theoretical input on $f_P$ can allow a determination of the CKM
element.  The importance of measuring $\Gamma(P\to \ell\nu)$
depends on the particle being considered. For the $B$ system, $f_B$
is crucial for using measurements of $B^0$-$\overline{B}^0$ mixing
to extract information on the fundamental CKM parameters.  Knowledge
of $f_{B_s}$ is also needed, but this parameter cannot be directly
measured as the $B_s$ is neutral, so the violation of the SU(3)
relation $f_{B_s} = f_B$ must be estimated theoretically. This
difficulty does not occur for $D$ mesons as both the $D^+$ and
$D_s^+$ are charged, allowing the direct measurement of SU(3)
breaking and a direct comparison with theory. (In this note mention
of specific particle charge also implies the use of the
charge-conjugate partner.)

For $B^-$ and $D_s^+$ decays, the existence of a charged Higgs boson
(or any other charged object beyond the Standard Model) would modify
the decay rates; however, this would not necessarily be true for the
$D^+$ \cite{Hou,Akeroyd}.  More generally, the ratio of $\mu \nu$ to
$\tau \nu$ decays can serve as one probe of lepton universality
\cite{Hou,Hewett}.

As $|V_{ud}|$ has been quite accurately measured in super-allowed
$\beta$ decays \cite{Vud}, with a value of 0.97418(26), measurements
of $\Gamma(\pi^+ \to \mu^+{\nu})$ yield a value for $f_{\pi}$.
Similarly, $|V_{us}|$ has been well measured in semileptonic kaon
decays, so a value for $f_{K}$ from $\Gamma(K^- \to \mu^-
\bar{\nu})$ can be compared to theoretical calculations. Recently,
however, lattice gauge theory calculations have been claimed to be
very accurate in determining $f_K$, and these have been used to
predict $|V_{us}|$ \cite{Jutt}.

Next we review current measurements, starting with the charm system.
The CLEO collaboration has measured the branching fraction for
$D^+\to\mu^+\nu$ and recently updated their published result
\cite{fD}. By using the well measured $D^+$ lifetime of 1.040(7) ps
and assuming $|V_{cd}|=|V_{us}|=0.2255(19)$ \cite{BM}, they report
\begin{equation}
f_{D^+}=(205.8\pm 8.5\pm 2.5)~{\rm MeV}~.
\end{equation}
This result includes a 1\% correction for the radiative
$\mu^+\nu\gamma$ final state based on the estimate by Dobrescu and
Kronfeld \cite{Kron}.

Before we compare this result with theoretical predictions, we
discuss the $D_s^+$. Measurements of $f_{D_s}$ have been made by
several groups and are listed in Table~\ref{tab:fDs}
\cite{CLEO-c,CLEO-CSP,Belle-munu,CLEO,BEAT,ALEPH,L3,OPAL,Babar-munu}.
Early measurements actually determined the ratio of the leptonic
decay to some hadronic decay, usually
$\Gamma(D_s^+\to\ell^+\nu)/\Gamma(D_s^+\to\phi\pi^+)$. This
introduces a large additional source of error since the denominator
is not well known.  CLEO \cite{CLEO-c} has published absolute
branching fractions for $\mu^+\nu$ and $\tau^+\nu$,
$\tau^+\to\pi^+\overline{\nu}$, and in a separate paper
\cite{CLEO-CSP} for $\tau^+\nu$, $\tau^+\to e^+\nu\overline{\nu}$;
there is also an as-yet-unpublished result from Belle
\cite{Belle-munu} for $\mu^+\nu$.

\begin{table}[htb]
\begin{center}

\caption{Experimental results for ${\cal{B}}(D_s^+\to \mu^+\nu)$,
${\cal{B}}(D_s^+\to \tau^+\nu)$, and $f_{D_s^+}$. Numbers have been
updated using the $D_s^+$ lifetime of 0.50 ps. Results listed below the
``Average'' line have not been used in our average. The assumed value
of ${\cal{B}}_{\phi\pi}\equiv {\cal{B}}(D_s^+\to\phi\pi^+)$ is
listed if evident. ALEPH averages their two results to obtain a
value for $f_{D_s^+}$. \label{tab:fDs}}
\begin{tabular}{llccc}\hline\hline
Exp. & Mode  & ${\cal{B}}$& ${\cal{B}}_{\phi\pi}$ (\%) & $f_{D_s^+}$ (MeV) \\
\hline
CLEO-c  \cite{CLEO-c}& $\mu^+\nu$& $(5.94\pm 0.66\pm 0.31)\cdot 10^{-3}$ & & $264\pm 15\pm 7$\\
CLEO-c  \cite{CLEO-c}& $\tau^+\nu$  & $(8.0\pm 1.3\pm 0.4)\cdot 10^{-2}$&& $310\pm 25 \pm 8 $ \\
CLEO-c  \cite{CLEO-CSP}& $\tau^+\nu$ & $(6.17\pm 0.71\pm 0.36)\cdot 10^{-2}$&& $273\pm 16 \pm 8 $ \\
CLEO-c & combined & -- & & $274\pm 10\pm 5$\\
Belle \cite{Belle-munu}  & $\mu^+\nu$ & $(6.44\pm 0.76\pm 0.52)\cdot 10^{-3}$&& $275\pm 16 \pm 12 $ \\
\hline Average& &&&$275\pm 10$\\
\multicolumn{2}{l}{Average with radiative correction} & & & $273\pm 10$\\
 \hline
 CLEO \cite{CLEO}& $\mu^+\nu$ &$(6.2\pm 0.8\pm 1.3 \pm
1.6)\cdot 10^{-3}$&
3.6$\pm$0.9&$273\pm19\pm27\pm33$\\
BEATRICE \cite{BEAT} & $\mu^+\nu$ &$(8.3\pm 2.3\pm 0.6 \pm
2.1)\cdot 10^{-3}$& 3.6$\pm$0.9&$312\pm43\pm12 \pm39$\\
ALEPH \cite{ALEPH}& $\mu^+\nu$ &$(6.8\pm 1.1\pm 1.8)\cdot 10^{-3}$ & 3.6$\pm$0.9& $282\pm 19\pm 40$ \\
ALEPH \cite{ALEPH}& $\tau^+\nu$ &$(5.8\pm 0.8\pm 1.8)\cdot 10^{-2}$ & &  \\
L3 \cite{L3} &$\tau^+\nu$ & $(7.4\pm 2.8 \pm 1.6\pm 1.8)\cdot 10^{-2}$ & & $299\pm 57\pm 32 \pm 37$  \\
OPAL \cite{OPAL} & $\tau^+\nu$ & $(7.0\pm 2.1 \pm 2.0)\cdot 10^{-2}$ & & $283\pm 44\pm 41$  \\
BaBar \cite{Babar-munu} & $\mu^+\nu$&
(6.74$\pm$0.83$\pm$0.26$\pm$0.66) $\cdot 10^{-3}$ & 4.71$\pm$0.46 &
 $283\pm 17 \pm 7 \pm 14$\\\hline\hline
\end{tabular}
\end{center}
\end{table}

\begin{table}[htb]
\begin{center}
\caption{Theoretical predictions of $f_{D^+_s}$, $f_{D^+}$, and
$f_{D_s^+}/f_{D^+}$. QL indicates quenched lattice calculations.
(Only selected results with quoted errors are included.)}
\label{tab:Models}
\begin{tabular}{lccl} \hline\hline
    Model &$f_{D_s^+}$ (MeV) &  $f_{D^+}$ (MeV)          &  ~~~~~$f_{D_s^+}/f_{D^+}$           \\\hline
Experiment (our averages)  & $273\pm 10$ & $205.8 \pm 8.9$ & $1.33\pm 0.07$\\

 Lattice (HPQCD+UKQCD) \cite{Lat:Foll} & $241\pm 3$ & $208\pm 4$ &
$1.162\pm 0.009$\\
Lattice (FNAL+MILC+HPQCD) \cite{Lat:Milc} &
$249 \pm 3 \pm 16 $ & $201\pm 3 \pm 17 $&$1.24\pm 0.01\pm 0.07$ \\
QL (QCDSF) \cite{QCDSF} &
220$\pm$6$\pm$5$\pm$11 &206$\pm$6$\pm$3$\pm$22 &$1.07\pm 0.02\pm 0.02$ \\
 QL (Taiwan) \cite{Lat:Taiwan} &
$266\pm 10 \pm 18$ &$235 \pm 8\pm 14 $&$1.13\pm 0.03\pm 0.05$ \\
QL (UKQCD) \cite{Lat:UKQCD}&$236\pm 8^{+17}_{-14}$ & $210\pm 10^{+17}_{-16}$ & $1.13\pm 0.02^{+0.04}_{-0.02}$\\
QL \cite{Lat:Damir} & $231\pm 12^{+6}_{-1}$&$211\pm 14^{+2}_{-12}$ &
$1.10\pm 0.02$\\
QCD Sum Rules \cite{Bordes} & $205\pm 22$ & $177\pm 21$ & $1.16\pm
0.01\pm 0.03$\\
QCD Sum Rules \cite{Chiral} & $235\pm 24$&$203\pm 20$ & $1.15\pm 0.04$ \\
Field Correlators \cite{Field} & $260\pm 10$&$210\pm 10$ & $1.24\pm 0.03$ \\
Isospin Splittings \cite{Isospin} & & $262\pm 29$ & \\
\hline\hline
\end{tabular}
\end{center}
\end{table}

We extract the decay constant from the measured branching ratios
using $|V_{cs}|=0.9742$, and a $D_s$ lifetime of 0.50 ps. Our
experimental average,
\begin{equation}
f_{D_s}=(273\pm 10){\rm ~MeV},
\end{equation}
uses only those results that are absolutely normalized
\cite{reason}. We note that the experiments do not correct
explicitly for any $\ell^+\nu\gamma$ that may have been included. We
have included the radiative correction of 1\% in the $\mu^+\nu$
rates \cite{Kron}~(the $\tau^+\nu$ rates need not be corrected).
Other theoretical calculations show that this rate is a factor of
40--100 below the $\mu^+\nu$ rate for charm \cite{theories-rad}.

Table~\ref{tab:Models} compares the experimental $f_{D_s}$ with
theoretical calculations
\cite{Lat:Foll,Lat:Milc,QCDSF,Lat:Taiwan,Lat:UKQCD,Lat:Damir,Bordes,Chiral,Field,Isospin}.
While most theories give values lower than the $f_{D_s}$
measurement, the errors are sufficiently large, in most cases, to
declare success. A recent unquenched lattice calculation
\cite{Lat:Foll}, however, differs by more than three standard
deviations \cite{Crit-follana}. Remarkably it agrees with $f_{D^+}$
and consequently disagrees in the ratio $f_{D_s^+}/f_{D^+}$.

Akeroyd and Chen \cite{AkeroydC} first pointed out that leptonic
decay widths are modified by new physics.  Specifically, for the
$D^+$ and $D^+_s$, in the case of the two-Higgs doublet model
(2HDM), Eq.~(\ref{equ_rate}) is modified by a factor $r_q$
multiplying the right-hand side:


\begin{equation}
r_q=\left[1+\left(1\over{m_c+m_q}\right)\left({M_{D_q}\over
M_{H^+}}\right)^2 \left(m_c-m_q\tan^2\beta\right)\right]^2,
\end{equation}
where $m_{H^+}$ is the charged Higgs mass, $M_{D_q}$ is the mass of
the $D$ meson (containing the light quark $q$), $m_c$ is the charm
quark mass, $m_q$ is the light-quark mass, and $\tan\beta$ is the
ratio of the vacuum expectation values of the two Higgs doublets.
(Here we modified the original formula to take into account the
charm quark coupling \cite{KronPC}.) For the $D^+$, $m_d \ll m_c$,
and the change due to the $H^+$ is very small. For the $D_s^+$,
however, the effect can be substantial. One major concern is that we
need to know the value of $f_{D_s^+}$ in the Standard Model (SM). We
can take that from a theoretical model. Our most aggressive choice
is that of the unquenched lattice calculation \cite{Lat:Foll},
because they claim the smallest error. Since the charged Higgs would
lower the rate compared to the SM, in principle, experiment gives a
lower limit on the charged Higgs mass. However, the value for the
predicted decay constant using this model is more than 3 standard
deviations {\it below} the measurement, implying that (a) either the
model of Ref.~\cite{Lat:Foll} is not representative; or (b) no value
of $m_{H^+}$ in the two-Higgs doublet model will satisfy the
constraint at 99.9\% confidence level; or (c) there is new physics,
different from the 2HDM, that interferes constructively with the SM
amplitude \cite{Rviolating}.

Dobrescu and Kronfeld \cite{Kron} emphasize that the discrepancy
between the theoretical lattice calculation and the CLEO data are
substantial and ``is worth interpreting in terms of new physics.''
They give three possible examples of new physics models that might
be responsible. These include a specific two-Higgs doublet model and
two leptoquark models.

The Belle \cite{Belle-taunu} and BaBar \cite{Babar-taunu}
collaborations have found evidence for $B^-\to\tau^-\bar{\nu}$
decays. The measurements are
\begin{eqnarray}
{\cal{B}}(B^-\to
\tau^-\overline{\nu})&=&(1.79^{+0.56~+0.46}_{-0.49~-0.51})\times
10^{-4}~~{\rm Belle};\\\nonumber &=&(1.2\pm0.4\pm0.3\pm0.2)\times
10^{-4}~~{\rm BaBar};\\\nonumber
&=&(1.42\pm0.43)\times 10^{-4}~~{\rm Our~average}.\\
\end{eqnarray}
The Belle and BaBar values have 3.5 and 2.6 standard-deviation
significances. More data are needed, and the average can only be
provisional. Here the effect of a charged Higgs is different as it
can either increase or decrease the expected SM branching ratio. The
factor $r$ is given in terms of the $B$ meson mass, $M_B$, by
\cite{Hou}
\begin{equation}
\label{eq:Hou} r=\left(1-\tan^2\beta\ {M_B^2\over m^2_{H^+}}\right)^2.
\end{equation}

In principle, we can get a limit in the $\tan\beta$--$m_{H^+}$ plane
even with this statistically limited set of data. Again, we need to
know the SM prediction of this decay rate. We ascertain this value
using Eq.~(\ref{equ_rate}). Here theory provides a value of
$f_B=(216\pm 22)$ MeV \cite{fBl}. The subject of the value of
$|V_{ub}|$ is addressed elsewhere \cite{HFAG}. Taking an average
over inclusive and exclusive determinations, and enlarging the error
using the PDG prescription because the results differ, we find
$|V_{ub}|=(3.9\pm0.5)\times 10^{-3}$, where the error is dominantly
theoretical. We thus arrive at the SM prediction for the
$\tau^-\bar{\nu}$ branching fraction of $(1.25\pm 0.41)\times
10^{-4}$. Taking the ratio of the experimental value to the
predicted branching ratio at its 90\% c.l.\ {\it upper} limit and
using Eq.~(\ref{eq:Hou}), we find that we can limit $M_{H^+}~/\tan
\beta > 3.5$ GeV. The 90\% c.l.\ {\it lower} limit also permits us
to exclude the region 4.1 GeV $< M_{H^+}~/\tan \beta < 9.6$ GeV
\cite{IsiPar}.

We now discuss the determination of charged pion and kaon decay
constants. The sum of branching fractions for $\pi^- \to \mu^- \bar
\nu$ and $\pi^- \to \mu^- \bar \nu \gamma$ is 99.98770(4)\%.  The
two modes are difficult to separate experimentally, so we use this
sum, with Eq.~(\ref{equ_rate}) modified to include photon emission
and radiative corrections \cite{Marciano-Sirlin}.  The branching
fraction together with the lifetime 26.033(5) ns gives
\begin{equation}
 f_{\pi^-} = (130.4\pm 0.04\pm 0.2)~{\rm MeV}~.
\end{equation}
The first error is due to the error on $|V_{ud}|$,
0.97418(26) \cite{Vud}; the second is due to the higher-order
corrections.

Similarly, the sum of branching fractions for $K^- \to \mu^- \bar
\nu$ and $K^- \to \mu^- \bar \nu \gamma$ is 63.57(11)\%, and the
lifetime is 12.3840(193) ns \cite{Antonelli}. We use a value for
$f_+(0)|V_{us}|$ obtained from the average of semileptonic kaon
decays of 0.21661(46). The $f_+(0)$ must be determined
theoretically. We follow Blucher and Marciano \cite{BM} in using the
Leutwyler-Roos calculation $f_+(0)=0.961\pm 0.008$, that gives
$|V_{us}|=0.2255\pm 0.0019$, yielding
\begin{equation}
f_{K^-} =(155.5 \pm 0.2\pm 0.8\pm 0.2)~{\rm MeV}~.
\end{equation}
The first error is due to the error on $\Gamma$; the
second is due to the CKM factor $|V_{us}|$, and the third is due to
the higher-order corrections. The largest source of error in these
corrections depends on the QCD part, which is based on one
calculation in the large $N_c$ framework.  We have doubled the
quoted error here; this would probably be unnecessary if other
calculations were to come to similar conclusions.  A large part of
the additional uncertainty vanishes in the ratio of the $K^-$ and
$\pi^-$ decay constants, which is
\begin{equation}
f_{K^-}/f_{\pi^-} = 1.193 \pm 0.002 \pm 0.006 \pm 0.001~.
\end{equation}
The first error is due to the measured decay rates; the
second is due to the uncertainties on the CKM factors; the third is
due to the uncertainties in the radiative correction ratio.

These measurements have been used in conjunction with lattice
calculations that predict $f_K/f_{\pi}$ in order to find a value for
$|V_{us}|/|V_{ud}|$.  Together with the precisely measured
$|V_{ud}|$, this gives an independent measure of $|V_{us}|$
\cite{Jutt,Antonelli}.

This work was supported by the U. S. National Science Foundation. We thank Bostjan Golob, William Marciano,
Soeren Prell and Charles Wohl for useful comments.


\begin{thebibliography}{99}
\bibitem{Bradcor}
Most predictions for the rate for $B^-\to\mu^-\bar{\nu}\gamma$
are in the range of 1 to 20 times the rate for
$B^-\to\mu^-\bar{\nu}$.
See G. Burdman, T. Goldman, and D. Wyler, \prD51,111(1995);
P. Colangelo, F. De Fazio, and G. Nardulli, \plB372,331(1996);
\ibid, {\bf 386}, 328 (1996);
A. Khodjamirian, G. Stoll, and D. Wyler, \plB358,129(1995);
G. Eilam, I. Halperin, and R. Mendel, \plB361,137(1995);
D. Atwood, G. Eilam, and A. Soni, \mplA11,1061(1996);
C.Q. Geng and C.C. Lih, \prD57,5697(1998) and
\mplA15,2087(2000);
G.P. Korchemsky, D. Pirjol, and T.M. Yang, \prD61,114510(2000);
C.W. Hwang, \epjC46,379(2006).


\bibitem{Hou}
W.-S. Hou, \prD48,2342(1993).


\bibitem{Akeroyd} See, for example, A.G. Akeroyd and S. Recksiegel,
\plB554,38(2003); A.G. Akeroyd, Prog.\ Theor.\ Phys.\ {\bf 111}, 295 (2004).


\bibitem{Hewett} J.L. Hewett, {\tt hep-ph/9505246}\rm,
presented at {\it Lafex
International School on High Energy Physics (LISHEP95)}\rm, Rio de Janeiro,
Brazil, Feb. 6-22, 1995.


\bibitem{Vud}
I. S. Towner and J. C. Hardy, \prC77,025501(2008)


\bibitem{Jutt}
W.J. Marciano, \prl93,231803(2004);
A. J\"uttner, {\tt [arXiv:0711.1239]\rm} (2007).


\bibitem{fD} S. Stone, ``Leptonic Decays: Measurements of $f_{D^+}$
and $f_{D_s}$,'' presented at FPCP 2008, Taipei, Taiwan, May, 2008,
to appear in the proceedings; we have included the radiative
correction here. See also M. Artuso \etal, (CLEO Collab.),
\prl95,251801(2005)

\bibitem{BM} CLEO assumes that $|V_{cd}|$ equals $|V_{us}|$ and takes the
value from E. Blucher and W. J. Marciano, ``$V_{ud}$, $V_{us}$, The
Cabibbo Angle and CKM Unitarity'' in PDG 2008.

\bibitem{Kron} B. A. Dobrescu and A. S. Kronfeld, arXiv:0803.0512
[hep-ph].

\bibitem{CLEO-c} M. Artuso \etal, (CLEO Collab.), \prl99,071802(2007).

\bibitem{CLEO-CSP}
K.M. Ecklund \etal, (CLEO Collab.), \prl100,161801(2008).


\bibitem{Belle-munu} K. Abe \etal, (Belle Collab.),
{\tt [arXiv:0709.1340]}\rm ~(2007).


\bibitem{CLEO}
M. Chadha \etal, (CLEO Collab.), \prD58,032002(1998).


\bibitem{BEAT}
Y. Alexandrov \etal, (BEATRICE Collab.), \plB478,31(2000).


\bibitem{ALEPH}
A. Heister \etal, (ALEPH Collab.), \plB528,1(2002).


\bibitem{L3}
M. Acciarri \etal, (L3 Collab.), \plB396,327(1997).


\bibitem{OPAL}
G. Abbiendi \etal, (OPAL Collab.), \plB516,236(2001).


\bibitem{Babar-munu}
B. Aubert \etal, (BABAR Collab.), \prl98,141801(2007).


\bibitem{reason}
We exclude measurements with normalizations to
other $D_s^+$ decay modes, mostly $\phi\pi^+$, because of the large
systematic uncertainty that this introduces. Furthermore,
due to the complications caused by interferences of resonances in
the $K^+K^-\pi^+$ Dalitz plot, the value of
${\cal{B}}(D_s^+\to\phi\pi^+)$ depends on the experimental
resolution and the width of the $K^+K^-$ mass interval used by each
experiment; see J. Alexander \etal~(CLEO Collaboration),
{\tt [arXiv:0801.0680]}\rm (2007).


\bibitem{theories-rad}
See most papers in Ref.~\cite{Bradcor} and C. D. L\"u and G. L. Song,
\plB562,75(2003).



\bibitem{Lat:Foll} E. Follana \etal, (HPQCD and UKQCD Collabs.),
 {\tt [arXiv:0706.1726]}\rm (2007). The statistical error is
 given as 0.6\%, with the other systematic errors added in
 quadrature yielding the 3 MeV total error.


\bibitem{Lat:Milc}
 C. Aubin \etal, (MILC Collaboration), \prl95,122002(2005).


\bibitem{QCDSF}
A. Ali Khan \etal, (QCDSF Collaboration), \plB652,150(2007).


 \bibitem{Lat:Taiwan}
 T.W. Chiu \etal, \plB624,31(2005).


\bibitem{Lat:UKQCD}
L. Lellouch and C.-J. Lin (UKQCD Collaboration), \prD64,094501(2001).


\bibitem{Lat:Damir}
D. Becirevic \etal, \prD60,074501(1999).


\bibitem{Bordes}
J. Bordes, J. Pe\~narrocha, and K. Schilcher, \jhep0511,014(2005).


\bibitem{Chiral}
S. Narison, {\tt [hep-ph/0202200]}\rm ~(2002).


\bibitem{Field} A.M. Badalian \etal, \prD75,116001(2007);
see also A.M. Badalian and B.L.G. Bakker, {\tt [hep-ph/0702229]}\rm ~(2007).


\bibitem{Isospin}
J. Amundson \etal, \prD47, 3059 (1993)



\bibitem{Crit-follana} The small errors quoted in Follana \etal
~\cite{Lat:Foll} are being discussed in the lattice community.  See,
for example, M. Della Morte {\tt [arXiv:0711.3160]}\rm ~(2007), page 6.


\bibitem{AkeroydC} A.G. Akeroyd and C.H.
Chen, \prD75,075004(2007) {\tt [hep-ph/0701078]}\rm.

\bibitem{KronPC} A. Kronfeld, private communication.

\bibitem{Rviolating}
An example of such a model is R-parity violating SUSY; see
A. G. Akeroyd and S. Recksiegel, \plB554,38(2003)
{\tt [hep-ph/0210376]}\rm.


\bibitem{Belle-taunu}
K. Ikado \etal, (Belle Collaboration), \prl97,251802(2006).


\bibitem{Babar-taunu}
We use the average of the two results from B. Aubert \etal, (BaBar Collaboration)
{\tt [arXiv:0705.1820]}\rm, and B. Aubert \etal, (BaBar Collaboration),
\prD76,052002(2007){\tt [arXiv:0708.2260]}\rm, presented in the
latter reference.


\bibitem{fBl}
A. Gray \etal, (HPQCD), \prl95,212001(2005).


\bibitem{HFAG}
Heavy Flavor Averaging Group: see
{\tt http://www.slac.stanford.edu/xorg/hfag/}\rm.


\bibitem{IsiPar}
In supersymmetric models there may be corrections which weaken our
bound; a slightly different formula for $r$ is given by  Isidori and
Paradisi, \plB639,499(2006) {\tt [hep-ph/0605012]}\rm
~that depends on a model dependent value of a parameter,
$\epsilon_0$.


\bibitem{Marciano-Sirlin}
W.J. Marciano and A. Sirlin, \prl71,3629(1993);
V. Cirigliano and I. Rosell, {\tt [arXiv:0707.4464v1]}\rm ~(2007).


\bibitem{Antonelli}M. Antonelli, "Precision Tests of Standard Model with
Leptonic and Semileptonic Kaon Decays," presented at XXIII Int.
Symp. on Lepton and Photon Interactions at High Energy, Daegu, Korea
(2007) arXiv:0712.0734 [hep-ex].


\end{thebibliography}
\end{document}